\documentclass[twocolumn]{aastex61}
\pdfoutput=1 
\usepackage{amsmath,amstext}
\usepackage[T1]{fontenc}
\usepackage{apjfonts} 
\usepackage[figure,figure*]{hypcap}
\usepackage{graphicx}
\usepackage{verbatim}

\newcommand{\Msun}{\ensuremath{\mathrm{M_{\Sun}}}} 

\shorttitle{PSR~J1411+2551: A New Low Mass DNS System}
\shortauthors{Martinez et al.}

\begin{document}

\title{Pulsar J1411+2551: A New Low Mass Double Neutron Star System}

\author{J. G. Martinez}
\affiliation{Max-Planck-Institut f\"{u}r Radioastronomie, Auf dem H\"{u}gel 69, D-53121 Bonn, Germany}
\author{K. Stovall}
\affiliation{National Radio Astronomy Observatory, P.O. Box 0, Socorro, NM 87801, USA}
\author{P. C. C. Freire}
\affiliation{Max-Planck-Institut f\"{u}r Radioastronomie, Auf dem H\"{u}gel 69, D-53121 Bonn, Germany}
\author{J. S. Deneva}
\affiliation{George Mason University, Resident at the Naval Research Laboratory, Washington, DC 20375, USA}
\author{T. M. Tauris}
\affiliation{Max-Planck-Institut f\"{u}r Radioastronomie, Auf dem H\"{u}gel 69, D-53121 Bonn, Germany}
\affiliation{Argelander-Institut f\"{u}r Astronomie, Universit\"{a}t Bonn, Auf dem H\"{u}gel 71, D-53121 Bonn, Germany}
\author{A. Ridolfi}
\affiliation{Max-Planck-Institut f\"{u}r Radioastronomie, Auf dem H\"{u}gel 69, D-53121 Bonn, Germany}
\author{N. Wex}
\affiliation{Max-Planck-Institut f\"{u}r Radioastronomie, Auf dem H\"{u}gel 69, D-53121 Bonn, Germany}
\author{F. A. Jenet}
\affiliation{Center for Advance Radio Astronomy, University of Texas at Rio Grande Valley, One West University Boulevard, Brownsville, TX 78520, USA}
\author{M. A. McLaughlin}
\affiliation{Department of Physics and Astronomy, West Virginia University, 111 White Hall, Morgantown, WV 26506, USA}
\author{M. Bagchi}
\affiliation{The Institute of Mathematics Science (IMSc-HBNI), 4th Cross Road, CIT Campus Taramani, Chennai 600 113, India}

\submitjournal{ApJ, in press}

\begin{abstract}
In this work, we report the discovery and characterization of PSR J1411+2551, a new binary pulsar discovered in the Arecibo 327 MHz Drift Pulsar Survey. Our timing observations of the radio pulsar in the system span a period of about 2.5 years. This timing campaign allowed a precise measurement of its spin period (62.4 ms) and its derivative (9.6 $\pm$ 0.7) $\times 10^{-20}\, \rm s\, s^{-1}$; from these, we derive a characteristic age of $> 9.1\,$Gyr and a surface magnetic field strength of $<$ 2.6 $\times 10^{9}$ G. These numbers indicate that this pulsar was mildly recycled by accretion of matter from the progenitor of the companion star. The system has an eccentric ($e\, = \, 0.17$) 2.61 day orbit. This eccentricity allows a highly significant measurement of the rate of advance of periastron, $\dot{\omega} = 0.07686 \pm 0.00046 ^{\circ}~{\rm yr}^{-1}$. Assuming general relativity accurately describes the orbital motion, this implies a total system mass M = $2.538 \pm 0.022 \, \Msun$. The minimum companion mass is $0.92\, \Msun$ and the maximum pulsar mass is $1.62\, \Msun$. The large companion mass and the orbital eccentricity suggest that PSR J1411+2551 is a double neutron star system; the lightest known to date including the DNS merger GW 170817. Furthermore, the relatively low orbital eccentricity and small proper motion limits suggest that the second supernova had a relatively small associated kick; this and the low system mass suggest that it was an ultra-stripped supernova.
\end{abstract}


\section{Introduction}
The first double neutron star (DNS) system, PSR B1913+16, was discovered in 1974 by \cite{ht75}. Continued timing of this system resulted in a measurement of the orbital decay due to the emission of gravitational waves as predicted by general relativity (GR). This was the first detection (albeit an indirect one) of gravitational waves, almost 40 years before LIGO's direct detection \citep{LIGO_1}.

Since then, 16 additional DNS systems have been discovered in the Galaxy \citep{Thomas_DNS_formation}, including one system in which both neutron stars (NSs) have been detected as radio pulsars, PSRs J0737$-$3039A and B \citep{0737A,doublePSR}. This system has an orbital period of only 2.4 hr and with the presence of two radio pulsed signals, it is a unique laboratory for tests of GR and alternative theories of gravity in the strong-field regime \citep{GRTestsdoublePSR}. The discovery of DNS systems stimulated the construction of ground-based interferometric detectors of gravitational wave sources and helped in statistical predictions of the collision rate of DNS systems \citep[e.g.][and references therein]{kim15}, many years prior to the recent detection of GW 170817 \citep{2017PhRvL.119p1101A}.

DNSs are fossils, preserving the endpoints of an exotic long journey of stellar evolution and binary interactions. 
By probing the distribution of NS masses, proper motions, orbital periods, and eccentricities of DNSs and even, in some cases, the misalignment angle between the NS and the orbital angular momentum (e.g., \citealt{2013ApJ...767...85F}), we can obtain crucial information from their past evolution; this provides important clues on the nature of supernovae and NS formation, as well as binary star interactions~\citep{Thomas_DNS_formation}.

These systems begin as binaries consisting of two massive main-sequence (MS) stars. In time, the more massive star will undergo dramatic envelope expansion 
and eventually a supernova (SN) explosion. This results in a system consisting of an NS and a high-mass MS companion.
As the companion evolves and extends its envelope, there will be a phase of mass transfer onto the NS; binaries in this evolutionary stage can be detected as high-mass X-ray binaries (HMXBs) \citep{XrayBinaries}. 
At this stage, the orbit is circularized by tidal forces. Eventually, following a common-envelope in-spiral and a new mass-transfer phase, the companion undergoes an SN explosion as well, forming a second NS \citep[see, e.g.][]{TaurisFormation, DuncReview}. DNS systems that have a relatively small measured eccentricity ($e < 0.2$) are presumed to have underwent an ultra-stripped SN with small ejecta mass and often small kicks \citep{Thomas_DNS_formation}.

In such a system, the older NS might be detected as a mildly recycled pulsar (spin periods of tens of milliseconds and relatively small magnetic field of $10^{9}-10^{10}$ G), which was spun up by accretion from the progenitor of the younger NS. The younger NS itself might be detected as a normal pulsar, with a high magnetic field ($10^{10} - 10^{13} \;{\rm G})$ and, in most cases, a much slower rotation. If the two NSs remain bound after the second SN, they form a DNS. The system's orbit will almost inevitably be eccentric, given the mass loss and the SN kick and the impossibility of tidal circularization after formation of two compact objects.

Despite their importance for tests of gravity theories and NS mass measurements, the 16 DNSs currently known represent a tiny fraction of the more than 2600 pulsars currently known \citep{atnf}. With such a small sample, the statistical properties of this population are still relatively poorly known. It is therefore important to find more of these systems. Finding new DNSs is among the top priorities of many recent pulsar surveys, like the
Arecibo ALFA pulsar survey \citep{Cordes2006}, Green Bank North Celestial Cap (GBNCC) survey \citep{Stovall2014}, the HTRU-N \citep{Barr2013} and HTRU-S \citep{Keith2010} surveys; the latter has been superseded by the SUPERB survey \citep{Keane2017}. All of these employ specialized acceleration search algorithms in an attempt to detect tight DNSs.

In this Letter, we report the discovery of PSR~J1411+2551, a 62 ms pulsar found in data from another of these surveys, the Arecibo 327 MHz Drift Pulsar Survey, or AO327 \citep{Deneva13}.  At the time of writing, this survey has discovered a total of 75 pulsars and transients, which include PSR~J2234+0611, a millisecond pulsar (MSP) with an eccentric orbit \citep{Deneva13,J2234_optical}. Among the transients discovered are several rotating radio transients, but thus far no fast radio bursts \citep{Deneva2016}. One of the pulsars found in this survey, PSR~J0453+1559, is a member of a DNS; this is currently the DNS with the largest mass asymmetry known \citep{J0453}. As we will see below, PSR~J1411+2551 is also a member of a DNS system; this is the second such system found by AO327.

The remainder of this paper follows as such. In section~\ref{sec:observations}, we describe the observations. In section~\ref{sec:timing}, we describe the derivation of the timing solution for this pulsar. We end with the conclusion (section~\ref{sec:discussion}) and a discussion of our findings.

\begin{figure}[h!]
\begin{center}
\includegraphics[width=2.3in, angle=270]{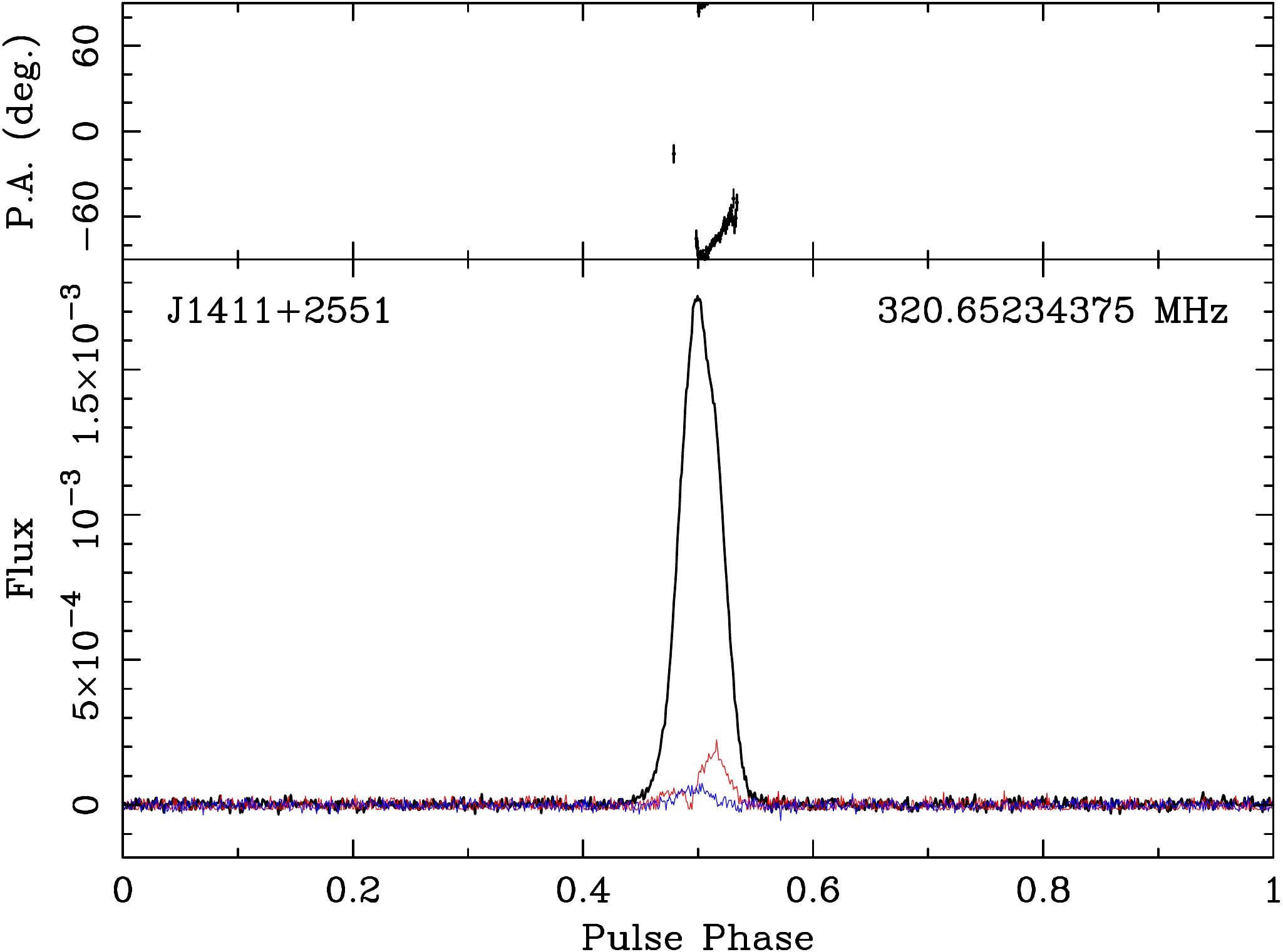}
\end{center}
\caption{Polarimetric pulse profile for PSR J1411+2551 at the frequency of 327 MHz, for a total bandwidth of 50 MHz; this was obtained by averaging the best detections of the pulsar at this frequency. The black line indicates total intensity in arbitrary units, the red line is the amplitude of linear polarization and the blue line is the amplitude of the circular polarization, all displayed as a function of spin phase. In the top panel we display the position angle of the linearly polarized component.}
\label{figure:profile}
\end{figure}

\section{Timing Observations}
\label{sec:observations}

PSR J1411+2551 was discovered on 2014 September 9 by one of us (KS) in data from the AO327 taken with the Puerto Rico Ultimate Pulsar Processing Instrument (PUPPI, a clone of the Green Bank Ultimate Pulsar Processing Instrument)\footnote{\url{http://safe.nrao.edu/wiki/bin/view/CICADA/GUPPISupportGuide} }. These observations were done in the incoherent search mode with a bandwidth of $\Delta f = 50$ MHz (from 302 to 352 MHz), 512 channels, a sampling time of 64 $\mu $s and two polarizations, with only the total power from these being recorded.    
After the discovery and until 2015 October 18, the pulsar was reobserved regularly
with the incoherent search mode with the same time and spectral resolution as in the discovery observation;
however, instead of just total intensity, we recorded the four Stokes parameters.

\begin{table}
\begin{center}{\footnotesize
\caption{PSR J1411+2551: Timing Solution}
\begin{tabular}{l c }
\hline
PSR & \multicolumn{1}{c}{J1411+2551}  \\
\hline \hline
Fitting program \dotfill                                       & TEMPO  \\
  Time units \dotfill                                          & TDB   \\
  Solar system ephemeris \dotfill                              & DE421  \\
  Reference epoch (MJD) \dotfill                               & 57617   \\
  Span of timing data (MJD) \dotfill                           & 57049-57915  \\
  Number of TOAs \dotfill                                      &  660  \\
  rms residual ($\mu s$) \dotfill                              & 32.77   \\
  Solar $n_0$ (cm$^{-3}$) \dotfill                             & 10.0   \\
  Right Ascension, $\alpha$ (J2000) \dotfill                   & 14:11:18.866(3)  \\
  Declination, $\delta$ (J2000) \dotfill                       & +25:51:08.39(7) \\
  Proper motion in $\alpha$, $\mu_{\alpha}$ (mas yr$^{-1}$) \dotfill                   & $-3$(12)  \\
  Proper motion in $\delta$, $\mu_{\delta}$ (mas yr$^{-1}$) \dotfill                   & $-4$(9)  \\
  Pulsar period, $P$ (s) \dotfill                            &  0.062452895517590(2) \\
  Period derivative, $\dot{P}$ ($10^{-20}$~ss$^{-1}$) \dotfill  &   9.56(51) \\
  Dispersion measure, DM (pc $cm^{-3}$ ) \dotfill              & 12.3737(3)   \\
  Rotation measure, (rad m$^{-2}$) \dotfill                     &  7.6(7) \\
  \hline
  \multicolumn{2}{c}{Binary Parameters}\\
  \hline\hline
  Orbital model \dotfill                                        & DD \\
  Orbital period, $P_b$ (days) \dotfill                         & 2.61585677939(8) \\
  Projected semi-major axis, $x$ (lt-s) \dotfill 				& 9.205135(2) \\
  Epoch of periastron, $T_0$ (MJD) \dotfill                     & 57617.04513(1)\\
  Orbital eccentricity, $e$ \dotfill                            & 0.1699308(4)  \\
  Longitude of periastron, $\omega$ ($^\circ$) \dotfill         & 81.5413(2)   \\
  Rate of advance of periastron, $\dot{\omega}$ ($^\circ \rm yr^{-1}$) \dotfill & 0.0768(4)   \\
  \hline
  \multicolumn{2}{c}{Derived parameters}\\
  \hline\hline
  Mass function, $f$ ($\Msun$ ) \dotfill                        & 0.1223898(9)   \\
  Total mass, $M$ ($\Msun$ ) \dotfill                           & 2.538(22)     \\
  Pulsar mass, $M_{p}$ ($\Msun$ ) \dotfill                      & $<$ 1.62     \\
  Companion mass, $M_c$ ($\Msun$ ) \dotfill                     & $>$ 0.92     \\
  Galactic longitude, $l$ \dotfill                              & 33.3789 \\
  Galactic latitude, $b$ \dotfill                               & 72.1009 \\
  DM derived distance, $d$ (kpc) \dotfill                       & 0.977 \\
  Galactic height, $z$ (kpc) \dotfill                           &  0.93  \\
  Surface magnetic field strength, $B_0$ ($10^{9}$ G) \dotfill & $\sim 1.8$ - 2.6 \\
  Characteristic age, $\tau_c$ (Gyr) \dotfill                   &  9.1 - $\sim$20   \\
  \hline
\multicolumn{2}{p{\columnwidth}}{Notes. The distance is derived from the DM using the \citet{CordesLazio2002} model of the Galactic electron density with a $\sim$25\% uncertainty. }
\end{tabular}
\vspace{-0.5cm}
}
\end{center}
\label{table:J1411+2551_timing_sol}
\end{table}

\begin{figure*}
 \centering
 \includegraphics[width=\textwidth,angle=0]{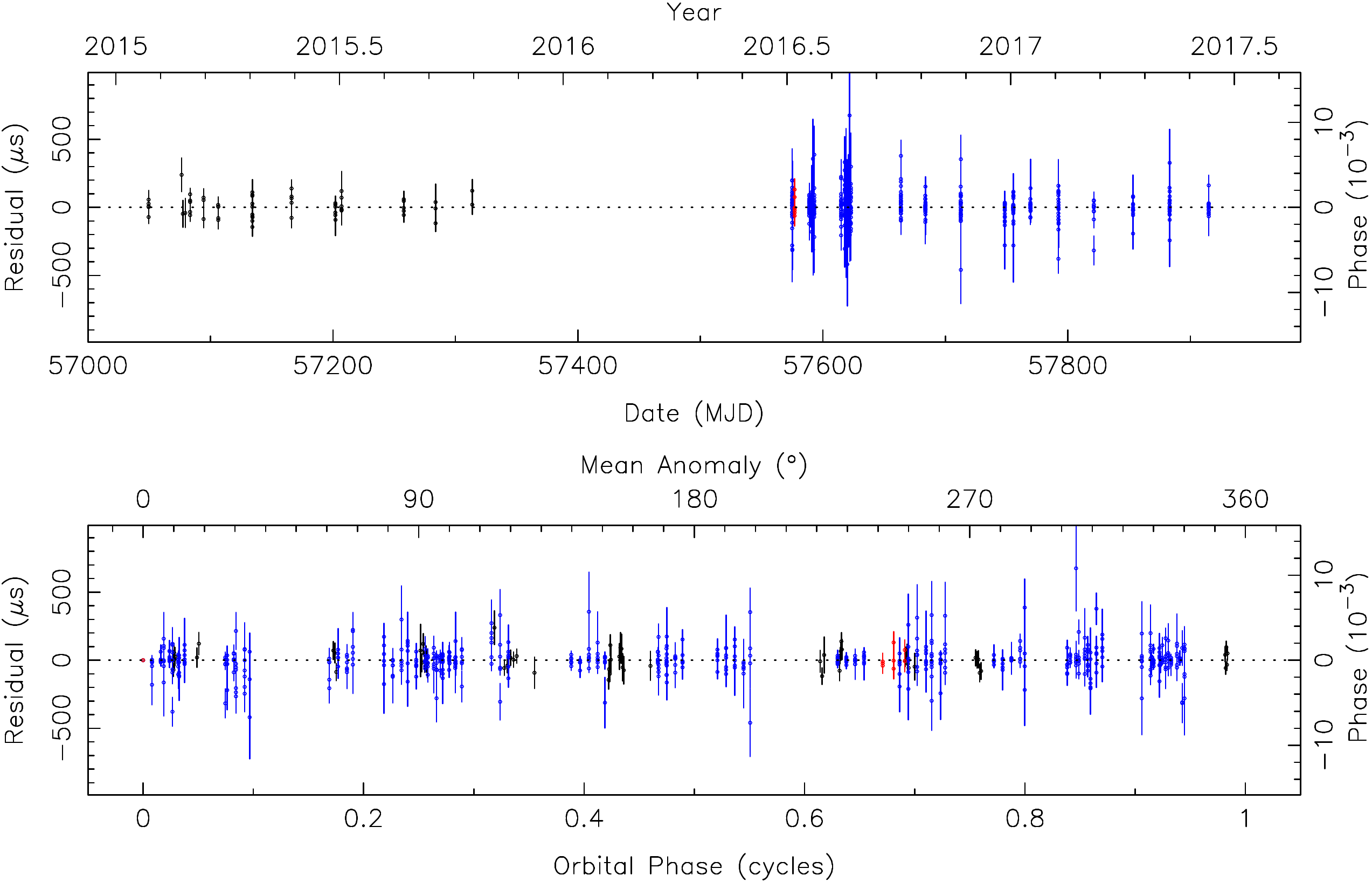}
 \caption{Post-fit residuals (time of arrival $-$ best-fit model) for the TOAs and timing solution of PSR~J1411+2551. {\em Top}: Residuals displayed versus epoch. {\em Bottom}: Residuals versus orbital phase.
 The black residuals correspond to the incoherent search mode data at 327 MHz, the blue ones to the coherently dedispersed data at 327 MHz and the red to the coherently dedispersed data at L band. All data were taken with PUPPI. No significant trends
 are detectable in the residuals, showing that the ephemeris in Table~\ref{table:J1411+2551_timing_sol}
provides a good description of the data.}
 \label{fig:residuals}
\end{figure*}

After a gap of several months, we continued observing the pulsar from 2016 July 5 until 2017 June 11. These observations were taken in the coherently dedispersed search mode, with a sampling time of 10~$\mu$s, as described by \cite{J0453}. Like the original search data, these data retain sensitivity to a possible pulsar signal from the companion NS and allow post-processing removal of radio frequency interference (RFI), but with much improved time resolution.

The subsequent analysis is also similar to that described in detail in
\cite{J0453}. First, we fold all the data to obtain pulse profiles. These are then calibrated using the noise diode observations taken before each observation. Each profile is then corrected for the Faraday effect, which requires the measurement of the rotation measure (RM). This is obtained by using the {\tt rmfit} routine of the {\tt PSRCHIVE} software \citep{PSRCHIVE,PSRCHIVE2012}, which looks for the best-fitting RM value by maximizing the linearly polarized intensity of the frequency-averaged profile. 
In Figure~\ref{figure:profile}, we show the result of the sum of the coherently dedispersed, Faraday-corrected profiles.

For each of the calibrated profiles, the individual pulse profiles are then cross-correlated with the template of Figure~\ref{figure:profile} using the procedure described in \cite{TaylorRG} and implemented the {\tt PSRCHIVE}. This produced 660 topocentric pulse times of arrival (TOAs). 
We then used {\tt TEMPO}\footnote{\url{http://tempo.sourceforge.net/} } to estimate the barycentric TOAs and then to estimate the pulsar model parameters by minimizing the root mean square (rms) of the timing residuals (calculated as the difference between the measured TOA and the model prediction for the same pulse). These parameters are presented in Table~\ref{table:J1411+2551_timing_sol}. We used the theory-independent DD binary model \citep{Damour85,Damour86} to describe the orbital motion.

The TOAs were split into two subsets. The first set consists of the incoherent search mode observations and the second set consists of are the coherently dedispersed search mode observations. Each subset of the TOA uncertainties was scaled using an EFAC parameter of 0.94 and 1.06 such that the reduced $\chi^2$ = 1 when each subset was fit independently of the others. The timing residuals are displayed in Figure~\ref{fig:residuals}. Their rms is 33 $\mu$s; this is a fraction of $5.2 \times 10^{-4}$ of the spin period. No significant trends are detectable in the residuals, showing that the ephemeris in Table~\ref{table:J1411+2551_timing_sol} provides a good description of the TOAs.

\section{Results}
\label{sec:timing}
The pulsar's ephemeris in Table~\ref{table:J1411+2551_timing_sol} includes a precise sky position, which allows for optical follow-up. No optical counterpart to the system is detectable in the online DSS2 optical survey, either in the red or blue filters, nor in the 2MASS survey.
 
The ephemeris also includes precise measurements of the pulsar's spin period ($P$) and its derivative ($\dot{P}$). From $\dot{P}$ we subtract the term due to the difference in Galactic acceleration between
the PSR J1411+2551 and the Solar System, $-1.3 \, \times \, 10^{-20} \rm s \, s^{-1}$ \citep{2017MNRAS.465...76M}, to derive 
a maximum intrinsic $\dot{P}_{\rm int} \, = \, 10.9 \, \times \, 10^{-20} \rm s \, s^{-1}$. From this, we derive a minimum characteristic age $\tau_c \, = \, 9.1\, \times \, 10^{9}$ years and a maximum surface inferred magnetic field of $B_0 \, = \, 2.6 \, \times \, 10^{9}$\;{\rm G}. These values can change significantly if we add the effects of the proper motion \citep{Shklovskii70}: for a proper motion of 20 mas yr$^{-1}$ (close to our upper limit), we would have 
$\dot{P}_{\rm int} \, = \, 4.9 \, \times \, 10^{-20} \rm s \, s^{-1}$, $\tau_c \, = \, 20\, \times \, 10^{9}\, \rm years,$ and
$B_0 \, = \, 1.8 \, \times \, 10^{9}$\;{\rm G}.
This
indicates that PSR~J1411+2551 was recycled by accretion of mass from the progenitor of its companion (\citealt{tlk12}).

From the orbital period $P_b$ (2.6 days) and the projected semi-major axis $x$ 
(9.20 light seconds), we obtain the mass function
\begin{equation}
f(M_{p},M_{c})  =  \frac{(M_{c} {\rm sin} i)^{3}}{ M^{2}}
                =  \frac{4\pi^{2}x^{3}}{T_{\odot}P_{b}^{2}} = 0.1223893(9)\, \Msun,
\label{eq:massf}
\end{equation}
where $T_{\odot} = G \Msun c^{-3} = 4.925490947 \mu \rm s$ is the solar mass times Newton's gravitational constant ($G \Msun$, a quantity known much more precisely than either $G$ or $\Msun$)
divided by the cube of the speed of light $c$, $i$ is the angle between the line of sight and the orbital angular momentum, and $M$ is the total mass of the system. This quantity and the
individual NS masses, $M_{p}$ and $M_{c}$, are here expressed in solar masses.

The orbital eccentricity of the system ($e = 0.17$) allows for a detection of one post-Keplerian (PK) parameter, the rate of advance of periastron ($\dot{\omega}$). If we assume this to be purely relativistic, then it depends only on the total mass of the system $M$ and Keplerian orbital parameters, which are already known precisely \citep{rob38,tw82}:
\begin{equation}
M\,=\,\frac{1}{T_{\odot}} \left[ \frac{\dot{\omega}}{3} (1- e^2)
  \right]^{\frac{3}{2}} \left( \frac{P_{\mathrm{b}}}{2\pi}
\right)^{\frac{5}{2}}.
\label{eq:M}
\end{equation}
Our measurement of $\dot{\omega}$ yields $M \, = \, 2.538 \, \pm \,  0.022 \, \Msun$.
This could be the lightest DNS known, but this is not clear yet: the previous
lightest DNS, PSR~J1756$-$2251, has a total mass of $2.56999(6)\, \Msun$ \citep{Ferdman14},
the difference in mass between the two systems, $0.032\, \Msun$, is only 1.45 times the 1$\sigma$
uncertainty on the mass of the PSR~J1411+2551 system.

From the total mass and the Keplerian mass function, we can obtain a lower limit for $M_c$ (assuming $i = 90^\circ$):
\begin{equation}
 M_{c} > \sqrt[3]{M^{2} f(M_{p},M_{c})} \approx 0.92 \Msun.
 \label{eq:massC}
\end{equation}
Given the value of $M$, this implies $M_{p}\, < \, 1.62\, \Msun $. The remaining PK parameters are, for the time being, not measurable in this system.

We have searched for radio pulsations from the companion in observations that were taken in the coherently dedispersed search mode. These observations were first dedispersed at the nominal DM of PSR J1411+2551. After that, we removed the orbital modulation due to the putative companion orbit with a code already used in \citet{J0453}, making the companion appear as if it were isolated, and thus maximizing our search sensitivity. This technique, however, requires the knowledge of the system mass ratio, $q = M_{\rm p} / M_{\rm c}$, which is unknown. Nevertheless, assuming a conservative minimum mass for an NS of $0.8~\Msun$, and using the total system mass and the limits on $M_{\rm p}$ and $M_{\rm c}$ reported above, we find that the value of $q$ must be between $\sim (0.46-1.76$). Hence, we treated $q$ as a free parameter and let it vary within this range, with a sensible choice of the step size. Each demodulated time series produced was then searched with the {\tt PRESTO} pulsar search code.\footnote{http://www.cv.nrao.edu/$\sim$sransom/presto/} No pulsations coming from the companion NS were found.

\subsection{Formation of the PSR J1411+2551 system}

\begin{figure}
 \centering
 \includegraphics[width=0.6\columnwidth,angle=-90]{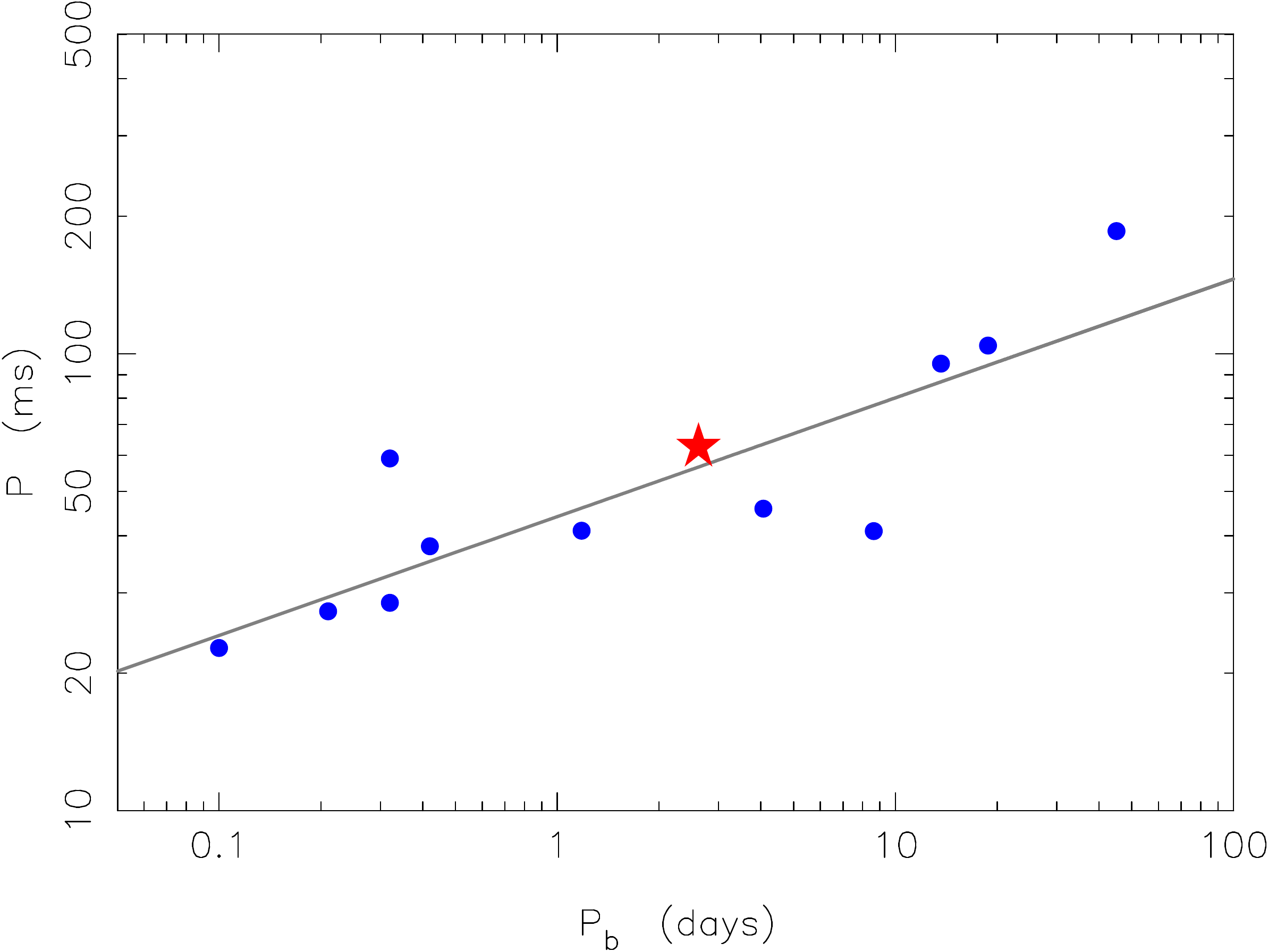}
 \caption{Blue points represent the spin period of the recycled pulsars in DNS systems as a function of their orbital period. PSR J1411+2551 is represented by the red star. The gray line represents Eq.~\ref{eq:PorbP}.
For a detailed discussion, see \cite{Thomas_DNS_formation}.}
 \label{fig:PorbP}
\end{figure}

Following the reasoning presented in the Introduction, the recycled nature of PSR~J1411+2551 and
its eccentric orbit, indicate that it is a member of a DNS system.
If the companion had slowly evolved into a massive white dwarf (WD), no supernova event
(with associated kick and mass loss) would occur, and the orbit would still be (nearly) circular.
The measured basic properties of the system ($P=62\;{\rm ms}$, $P_b=2.62\;{\rm days},$ and $e=0.17$) match very well that of other DNS systems. For example, the spin period of 62~ms fits nicely with the observed $(P_b,P)$ relation for recycled pulsars in similar systems \citep{Thomas_DNS_formation}:
\begin{equation}
 P\approx 44\;{\rm ms}\;\;(P_b{\rm /days})^{0.26}.
\label{eq:PorbP}
\end{equation}
This is seen in Figure~\ref{fig:PorbP}. Also the relatively small measured eccentricity of PSR~J1411+2551 is a typical value of the subgroup of DNS systems that are thought to have undergone an ultra-stripped SN with small ejecta mass and often small kicks \citep{tlm+13,tlp15,Thomas_DNS_formation}.
\\

We have simulated the kinematic effects of 200 million SN explosions in order to reproduce the measured orbital parameters of the PSR~J1411+2551 system, following the method applied in \citet{Thomas_DNS_formation}. 
As in that paper, in our Monte Carlo simulations, we assume that $M_{\rm c}$ falls within the range of masses measured for young NSs 
in other DNS systems ($1.17-1.39\;\Msun$) with a flat probability distribution. $M_p$ is then obtained from $M - M_c$.

The simulated solutions for the pre-SN binary show an enhanced probability for a small mass of the exploding star (i.e. $<2.5\;\Msun$) and a small associated kick of $<100\;{\rm km\,s}^{-1}$, as expected for ultra-stripped star progenitors (the kick value distribution peaks near $50\;{\rm km\,s}^{-1}$, although solutions are found up to almost $200\;{\rm km\,s}^{-1}$). The pre-SN orbital period was somewhere in the interval $1.4-3.3\;{\rm days}$.

Finally, we notice that the merger time of PSR~J1411+2551 is $\sim 460\;{\rm Gyr}$, given its relatively wide orbit. Therefore, despite its potentially large true age inferred from a small value of $\dot{P}$, there have been no changes in its orbital parameters at any significant level since its formation.

\begin{figure}[h]
\centering
\includegraphics[width=0.9\columnwidth,angle=0]{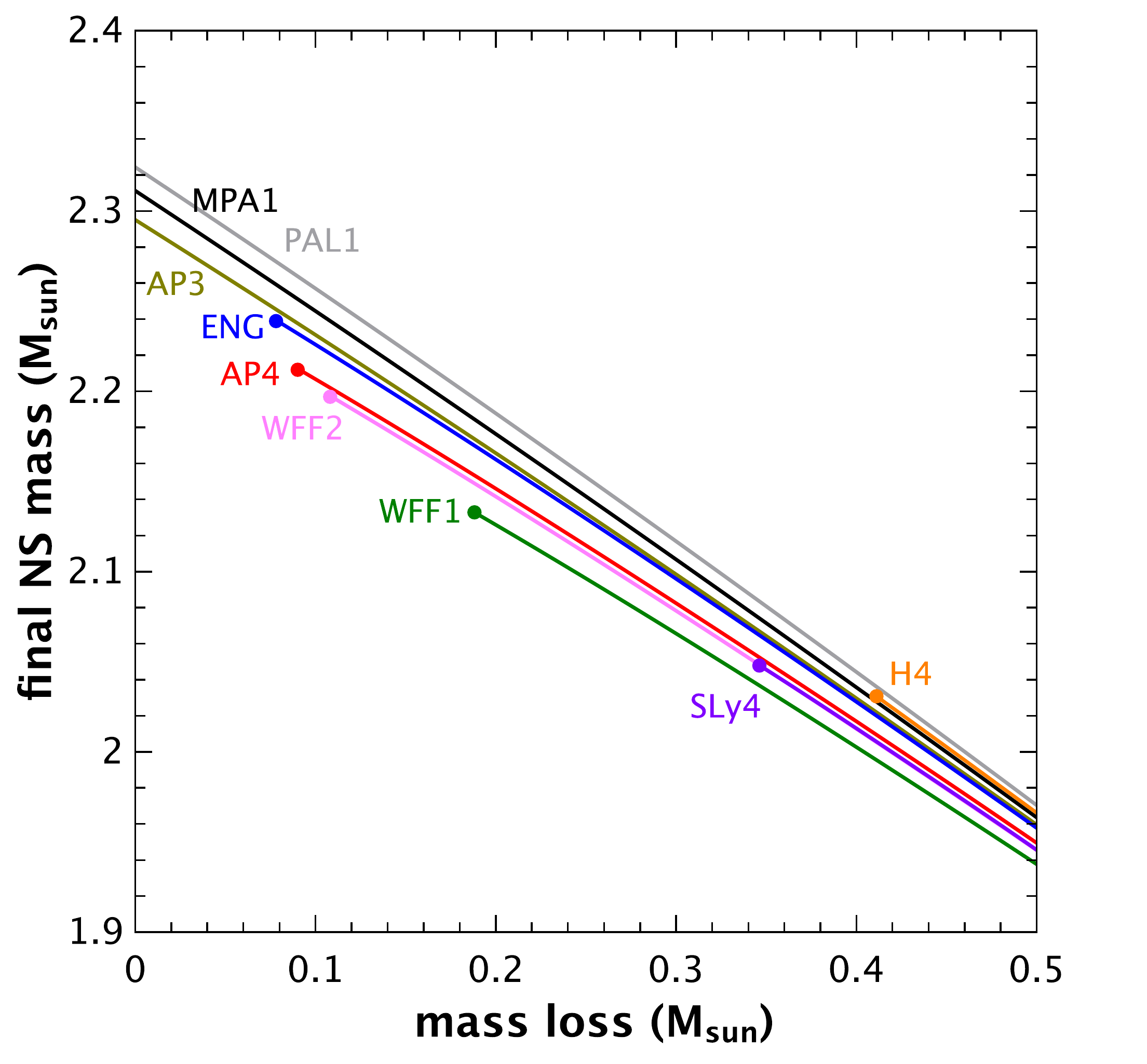}
\caption{Final NS mass for various EoS as a function of baryonic mass loss in a DNS merger event for a system similar to PSR ~J1411+2551. Only the AP3, MPA1, and PAL1 EoS models are able to leave behind a stable (slowly rotating) NS if the baryonic mass loss is less than about $0.05\;M_{\odot}$. For the different EoSs, see \cite{lp01}. Note, the EoS models H4 and PAL1 correspond to rather large neutron stars (typical radii $\sim$14\,km) and are therefore disfavored by the GW 170817 merger event \citep{2017PhRvL.119p1101A}.}
\label{fig:NS_or_BH}
\end{figure}

\section{Discussion and Conclusions}
\label{sec:discussion}

We have presented the discovery and timing solution of PSR J1411+2551, a 62.4 ms pulsar in an eccentric binary orbit with an NS found in the Arecibo 327 MHz Drift Pulsar Survey. We searched for the companion NS as a radio pulsar, but the search did not result in a detection. The 2.5 years of timing observations provided the detection of the rate of advance of periastron for the orbit. Assuming GR, it resulted in the measurement of a total mass for the system of M = 2.538 $\pm$ 0.022 \Msun, making it the lightest total mass measurement for a known DNS system to date. 

Because of the PSR J1411+2551 long orbital period, it is unlikely that
we will be able to measure other PK parameters. As in the case of J0453+1559 \citep{J0453}, the
Einstein delay ($\gamma$) will be strongly correlated with the kinematic $\dot{x}$ induced by the proper motion
(\citealt{Arzoumanian96} \& \citealt{Kopeokin96ProperMotion}). The variation of the orbital period caused by the emission of
gravitational waves ($\dot{P}_{\rm b, predicted} \, \simeq \, -6.24 \times 10^{-15} \rm s \, s^{-1}$ for nearly equal NS masses) will likely be undetectable given the likely much larger kinematic contributions: $\dot{P}_{\rm b, k} \, = \, -4.70 \times 10^{-14} \rm s \, s^{-1}$ from the Galactic acceleration \citep{2017MNRAS.465...76M}
and $\dot{P}_{\rm b, k} \, = \, 2.14 \times 10^{-13} \rm s \, s^{-1}$ for proper motions of up to $\sim \, 20\, \rm mas \, yr^{-1}$ \citep{Shklovskii70}. The Shapiro delay was not
detectable for this system; this implies that it does not have a high orbital inclination.
Without any other PK parameters, we cannot measure the individual masses of the two NSs in the system.

For investigating the destiny of merging DNS systems, it is of utmost importance to probe the possible final masses of the DNS merger products. If the merger product is less massive than some critical value then, rather than forming a black hole directly, it may provide a long-lived central engine in terms of a proto-magnetar (i.e. a fast rotating object emitting strong magneto-dipole radiation) that powers the extended X-ray emission observed in a large fraction of short gamma-ray bursts \citep[e.g.][and references therein]{zm01,mgt+11,rom+13,rk15}. However, the existence of such meta-stable NSs depends on both the NS equation of state (EoS) and the possible masses of the DNS merger products.

The DNS system J1411+2551 provides the smallest total mass ($2.54\;\Msun$) of any DNS system found so far, including the total mass estimated for the DNS merger event GW 170817 detected recently by the LIGO-Virgo collaboration \citep{2017PhRvL.119p1101A}. In Figure~\ref{fig:NS_or_BH}, we demonstrate that the resulting remnant mass of a DNS merger from an equivalent system would yield a total gravitational mass of about $2.30\;\Msun$ using some of the EoS from \cite{lp01}. This is even a conservative upper limit since an outflow of baryonic matter from the collision will remove up to $\sim 0.05\;M_{\odot}$ as inferred from GW 170817 \citep{dps+17}. 
Thus, the final compact object left behind is only some $0.25\;M_{\odot}$ heavier than the current record holder for a massive NS, PSR~J0348+0432, which has a mass of $2.01\pm0.04\;\Msun$ \citep{Antoniadis}. Note that, among the different EoS models we used, AP3, MPA1, and PAL1 are able to leave behind a stable NS in this case. This result would be interesting to compare with the constraints of EoS coming from NICER (Neutron star Interior Composition Explorer) and further LIGO--Virgo detections in the near future.

\acknowledgments
Acknowledgments: 
J.G.M. was supported for this research through a stipend from the International Max Planck Research School (IMPRS) for Astronomy and Astrophysics at the University of Bonn and Cologne and acknowledges partial support through the Bonn-Cologne Graduate School (BCGS) Honors Branch of Physics and Astronomy. P.C.C.F. and A.R. gratefully acknowledge financial support by the European Research Council for the ERC Starting grant BEACON under contract No. 279702, and continuing support from the Max Planck Society. J.S.D. was supported by the NASA Fermi program. K.S. was supported by the NANOGrav NSF Physics Frontiers Center award number 1430284.



\end{document}